\documentclass[aps,pra,showpacs,twocolumn]{revtex4}
\usepackage{amssymb}
\usepackage{epsfig}
\usepackage{graphicx}
\usepackage{amsmath}
\usepackage{array,color}
\usepackage{mathrsfs}
\usepackage{subfigure}
\begin{document}
\title{Fragmentation of Spin-orbit Coupled Spinor Bose-Einstein Condensates}
\author{Shu-Wei Song$^{1,2,3}$, Yi-Cai Zhang$^{4}$, Hong Zhao$^{1,2,3}$, Xuan Wang$^{1,2,3}$ and Wu-Ming Liu$^{4}$}

\address{$^1$State Key Laboratory Breeding Base of Dielectrics Engineering, Harbin University of Science and Technology, 150080 Harbin, China}
\address{$^2$Key Laboratory of Engineering Dielectrics and Its Application, Ministry of Education ,Harbin University of Science and Technology, 150080 Harbin, China}
\address{$^3$College of Electrical $\&$ Electronic Engineer, Harbin University of Science and Technology, Harbin, Heilongjiang, 150080, China}
\address{$^4$Beijing National Laboratory for Condensed Matter Physics,
Institute of Physics, Chinese Academy of Sciences, Beijing 100190, China}
\date{\today}

\begin{abstract}
The fragmentation of spin-orbit coupled spin-1 Bose gas with a weak interaction in external harmonic trap is explored by both exact diagonalization and mean-field theory. This fragmentation tendency, which originates from the total angular momentum conservation, is affected obviously by the spin-orbit coupling strength and the spin-dependent interaction. Strong spin-orbit interaction raises the inverse participation ratio, which describes the number of significantly occupied single-particle states. As the spin-dependent interaction changes from anti-ferromagnetic to ferromagnetic, the peak values in the inverse participation ratio become lower. Without the confinement of the appointed total angular momentum, the condensate chooses a zero or finite total angular momentum ground state, which is determined by both the interaction and the spin-orbit coupling strength.
\end{abstract}


\maketitle
\section{INTRODUCTION}
Since the experimental realization of the artificial external Abelian or non-Abelian gauge potentials coupled to neutral cold atoms, the spin-orbit coupling phenomena have been attracting extensive explorations \cite{YJLin2009, YJLin2011,YJLin2011natrue, MAidelsburger, ZFu2011, DLCampbell2011, Lawrence2012, PWang2012,zyc2013,lry2013,zhuguobao2013,zhangshangshun2013}, such as the studies on topics of Spin Hall effects \cite{MCBeeler2013}, Majorana fermions \cite{LiuXiaoJi2012,LiuXiaoJi2012_1}, etc.
In the presence of the magnetic field, the quantum Hall phases which occur in the vicinity of the degeneracy point was also studied \cite{TGrass2013}.
Due to the presence of the spin-orbit coupling mechanism, the energy spectrum changes dramatically. The single particle energy minimum has finite momentum and the ground states are circularly degenerate, in which case the ground state of the condensate favors mainly ``stripe" or ``plane wave" phase \cite{HuiZhai2012,ssw2013}. As pointed out in Refs. \cite{CJWu2011, Gopalakrishnan}, the two states locating at two opposite ends of the degenerate circle are free of exchange interactions, and the fragmented and coherent condensates in terms of this two states can be defined accordingly \cite{CJWu2011}; the ground state was found involving a fragmented condensate with respect to the two states in finite systems. With the interplay of the spin-orbit coupling and harmonic traps, the energy spectrum is very similar to the well-known Landau levels \cite{HuiHu2012}. Half vortex state or skyrmion lattice patterns emerge in the case of weakly interacting bosons.

To explore the formation and properties of vortices in an atomic boson system, the non-vanishing angular momentum states of weakly interacting Bose gase in harmonic trap have attracted considerable attentions \cite{TPapenbrock,GMKavoulakis,ADJackson, WJHuang2001,XJLiu2001}. In Ref \cite{XJLiu2001}, Xia-Ji Liu \emph{et al.} studied the ground state for a weakly interacting, harmonically trapped N-boson system and found that the ground state is generally a fragmented condensate because of the orbital angular momentum conservation. For the scalar Bose gas, fragmented states have been explored massively in Refs. \cite{GMKavoulakis, ADJackson2001,Wilkin1998,Bertsch1998}. Fragmented condensate also exists in a spin-$1$ Bose gas in uniform magnetic fields, which would be turned into a single condensate by magnetic field gradients \cite{Ho2000}.

\begin{figure}[htbp]
\centering
\includegraphics[width=9.40cm]{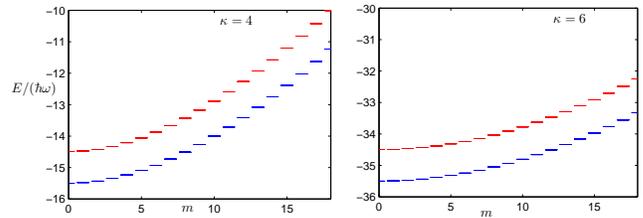}
\caption{The single-particle energy spectrum with total angular momentum good quantum number $m\geq0$ for spin-orbit coupled spin-$1$ bosons (the radial quantum numbers are $n=0$ for blue levels and $n=1$ for red levels). As the spin-orbit coupling strength $\kappa$ increases, the levels incline to resemble the well-known Landau levels spectrum structure.}
\label{SPSpectrum}
\end{figure}

As the spin-orbit coupling mechanism modifies the energy spectrum significantly (as shown in Fig. \ref{SPSpectrum}), what is the non-vanishing angular momentum state structure of the spin-orbit coupled spinor Bose gas in harmonic traps with a weak interaction? How does the variation of the spin-orbit coupling strength affect the quantum state structure? And what are the roles of the spin-dependent interaction parameters? As far as we know, these issues have not been addressed elsewhere.

In the present work, we explore the spin-orbit coupled spin-1 Bose gas with a weak interaction in the presence of the external harmonic trap. The fragmented condensate also arises at certain total angular momentums. As the spin-orbit coupling strength increases, this fragmentation tendency becomes more obvious. The spin-dependent interaction plays important roles in the particle distribution among the single-particle states with respect to the total angular momentum. When the interaction varies from anti-ferromagnetic to ferromagnetic, the fragmentation inclines to be suppressed. Thus the fragmentation is more favorable in anti-ferromagnetic Bose gas than that in ferromagnetic one. In addition, the ground state of the Bose gas is found to be with zero or finite total angular momentum.

The rest of the paper is organized as follows. In Sec. \ref{sec1}, the theoretical model is given, including the single-particle states and the second quantization form of the total Hamiltonian. Sec. \ref{sec2} is composed of the numerical results, including the roles of the spin-orbit coupling strength, spin-dependent interactions.
Sec. \ref{sec3} focuses on the ground state of the condensate without the restriction of appointed angular momentum.
Finally, the conclusions are summarized in Sec. \ref{sec4}.

\bigskip
\section{Energy Spectrum}\label{sec1}

In general, the constituent atoms in ultracold atomic gas have internal degrees of freedom originating from the spin. For atoms confined in an isotropic disk harmonic trap, the wave function for the z direction is frozen into the ground-state $f(z)=\mathrm{exp}(-z^2/za_{z}^2)/\pi^{1/4}a_{z}^{1/2}$ ($a_{z}=\sqrt{\hbar/M_{a}\omega_{z}}$, $M_{a}$ and $\omega_{z}$ are the atom mass and the confinement frequency in $z$ direction, respectively) if the interaction energy is sufficiently small compared to the energy spacing in the $z$ direction, i.e. $\hbar\omega_{z}$.  Thus, it is reasonable to consider the 2D isotropic harmonic trap. We consider a three-component Bose gas trapped in the 2D harmonic trap $V=M_{a}\omega^{2}(x^2+y^2)/2$ in the presence of the Rashba spin-orbit coupling effect: $V_{SO}=\gamma(\hat{p}_{x}S_{x}+\hat{p}_{y}S_{y})$, where $\omega$ is the confinement frequency in the $xy$ plane, $\gamma$ represents the strength of the spin-orbit coupling and $\hat{p}_{x,y}$, $S_{x,y}$ are the momentum operator and the spin-1 representation of Pauli matrices, respectively. The model Hamiltonian in the second quantization form is given by $\hat{H}= \hat{H}_0+ \hat{H}_{int}$ \cite{TinLunHo}, where
\begin{eqnarray}
\hat{H}_0&=&\int d\mathbf{r}\mathbf{\Psi}^\dag\left[\frac{\hat{\mathbf{p}}^2}{2M}
+\gamma(\hat{p}_xS_x+\hat{p}_yS_y)+{V}\right]\mathbf{\Psi},
\label{Hamiltonian1}\\
\!\!\hat{H}_{int}&=&\!\!\int\!\!d\mathbf{r}\left[\frac{c_0}{2}\!\psi^\dag_a \!\psi^\dag_{a'}\!\psi_{a'}\!\psi_{a}
+\!\!\frac{c_2}{2}\!\psi^\dag_a \!\psi^\dag_{a'}\!\mathbf{S}_{ab}\!\cdot\!\mathbf{S}_{a'b'}\!\psi_{b'} \!\psi_{b}\right].
\label{Hamiltonian2}
\end{eqnarray}
$\mathbf{\Psi}=(\psi_1,\psi_0,\psi_{-1})^T$ (the superscript $T$ stands for the transpose) denotes collectively the spinor
Bose field operators, and $\mathbf{S}=(S_x,S_y,S_z)$. In $\hat{H}_{int}$, the Einstein summation convention is used ($a,a'=1,0,-1$). The spin-independent and spin-dependent interactions are denoted as $c_0=4\pi\hbar^2(a_0+2a_2)/3M_{a}$ and $c_2=4\pi\hbar^2(a_2-a_0)/3M_{a}$, where $a_0$ and $a_2$ are the s-wave scattering lengths corresponding to the total spin of the two colliding bosons $0$ and $2$, respectively. The spin-orbit coupling parameter $\gamma$ is experimentally related to the wavelength of the laser beams and exact experimental setups. As proposed in reference \cite{PengjunWang}, $\gamma=2\pi\hbar \mathrm{sin}(\theta/2)/\lambda$, where $\theta$ and $\lambda$ are the angle between two Raman beams and the wavelength of the laser, respectively. In Ref. \cite{Gediminas}, the tetrapod setup used to generate the Rashba SO coupled BEC requires $\gamma={\sqrt{2}\pi \hbar}/{\lambda}$.

\subsection{Single-particle Energy Spectrum}\label{subsec11}
In the single-particle Hamiltonian, the spin-orbit coupling term reshapes the energy spectrum of the 2D harmonic oscillator, and levels that similar to the well-known Landau levels appear for sufficiently large spin-orbit coupling strength \cite{HuiHu2012}. For the 2D harmonic oscillator Hamiltonian $H_{Harm}=(\hat{p}_{x}^2+\hat{p}_{y}^2)/2M_{a}+M_{a}\omega^{2}(x^{2}+y^{2})$, the circular operators $\hat{a}_{d,g}$ can be introduced to manipulate the eigenstates and eigenvalues more conveniently. These circular orperators are $\hat{a}_{d}=(\hat{a}_{x}-i\hat{a}_{y})$ and $\hat{a}_{g}=(\hat{a}_{x}+i\hat{a}_{y})$, where $\hat{a}_{x}=(M_{a}\omega x+i\hat{p}_{x})/\sqrt{2M_{a}\hbar\omega}$ and $\hat{a}_{y}=(M_{a}\omega y+i\hat{p}_{y})/\sqrt{2M_{a}\hbar\omega}$ are destruction operators in the $Ox$ and $Oy$ directions, respectively. Thus, the 2D harmonic oscillator Hamiltonian can be written as $H_{Harm}=(\hat{a}_{d}^{\dagger}\hat{a}_{d}+\hat{a}_{g}^{\dagger}\hat{a}_{g}+1)\hbar\omega$. The corresponding eigenstates are \cite{Cohen-Tannoudji}

\begin{equation}
|\chi_{n_{d},n_{g}}\rangle=\frac{1}{\sqrt{n_{d}!n_{g}!}}(\hat{a}_{d}^{\dagger})^{n_{d}}(\hat{a}_{g}^{\dagger})^{n_{g}}|\varphi_{00}\rangle,
\end{equation}
where $|\varphi_{00}\rangle$ is the ground state of the two-dimensional harmonic oscillator.
In the polar coordinates $(\rho,\varphi)$, the eigenstates are
\begin{equation}
\chi_{mn}(\rho,\varphi)=R_{nm}(\rho)e^{im\varphi},
\end{equation}
where $n=n_{g}$, $m=n_{d}-n_{g}$, and $R_{nm}(\rho)=(-1)^{n}\sqrt{\frac{n!}{\pi(m+n)!}}e^{-\rho^{2}/2}L_{n}^{m}(\rho^{2})\rho^{m}$ ($L_{n}^{m}$ is the associated Laguerre polynomials).

The general wavefunction can be expanded in terms of the eigenfunctions: $|\chi_{n_{d},n_{g}}\rangle$,

\begin{equation}
|\psi\rangle=\sum_{n_{d},n_{g}}C_{n_{d}n_{g}}|\chi_{n_{d},n_{g}}\rangle, \label{Expansion}
\end{equation}
where $C_{n_{d}n_{g}}$ is the coefficients of the expansion.
For a state with certain angular momentum (quantum number $m$), it is only necessary to restrict the summation over $n_{d}$ and $n_{g}$ with $n_{d}-n_{g}=m$.

Because of the 2D isotropic harmonic potential, the single-particle wave-function corresponding to the single-particle Hamiltonian $H_{0}$ may have a well-defined azimuthal angular momentum. In fact, the Hamiltonian has rotational symmetries along z axial direction and the total angular momentum $J_{z}=L_{z}+S_{z}$ ($L_{z}$ and $S_{z}$ are the orbital and spin angular momentum, respectively) corresponds to a good quantum number. In polar coordinates $(\rho,\varphi)$, the eigenfunctions of the single-particle Hamiltonian $H_{0}$ can be written in the following form:
\begin{equation}
\Psi_{m}=\begin{pmatrix}\psi_{+}\\\psi_{0}\\\psi_{-}\end{pmatrix}=
\begin{pmatrix}\phi_{+}(\rho)e^{i(m-1)\varphi}\\\phi_{0}(\rho)e^{im\varphi}\\\phi_{-}(\rho)e^{i(m+1)\varphi}\end{pmatrix}\label{SingleParticleState}.
\end{equation}

In the energy spectrum seeking process, we restrict ourselves to $m\geq0$ because the eigenstates with $m<0$ can be obtained by using the time reversal symmetry. For particles with spin, the time reversal operator is
\begin{equation}
\Theta=\eta \mathrm{e}^{-i\pi \hat{S}_{y}/\hbar}K,
\end{equation}
where $\eta$ and $\hat{S}_{y}$ are arbitrary phase factor (can be conveniently chosen to be $1$) and the $y$ component of the spin operator, respectively. Here, $K$ stands for the complex conjugate operator. In the base kets composed of the $\hat{S}_{z}$ eigenkets, the time reversal matrix reads

\begin{equation}
\Theta=\begin{pmatrix}0&0&1\\0&-1&0\\1&0&0\end{pmatrix}K.
\end{equation}

By using the expansion in Eq. (\ref{Expansion}), we obtain
\begin{eqnarray}
|\psi_{+}\rangle&=&\sum_{n_{d},n_{g}}A_{n_{d}n_{g}}|\chi_{n_{d},n_{g}}\rangle, (n_{d}-n_{g}=m-1),\\
|\psi_{0}\rangle&=&\sum_{n_{d},n_{g}}B_{n_{d}n_{g}}|\chi_{n_{d},n_{g}}\rangle, (n_{d}-n_{g}=m),\\
|\psi_{-}\rangle&=&\sum_{n_{d},n_{g}}C_{n_{d}n_{g}}|\chi_{n_{d},n_{g}}\rangle, (n_{d}-n_{g}=m+1),
\end{eqnarray}
where $A_{n_{d}n_{g}}$, $B_{n_{d}n_{g}}$ and $C_{n_{d}n_{g}}$ are the corresponding coefficients of the expansions.

In terms of the circular operators $\hat{a}_{d,g}$, the single-particle Hamiltonian $\hat{H}_{0}$ reads
\begin{equation}
\hat{H}_{0}=(\hat{a}_{d}^{\dagger}\hat{a}_{d}+\hat{a}_{g}^{\dagger}\hat{a}_{g}+1)I
+\frac{\kappa}{\sqrt{2}}\mathcal{M}, \label{SingleParticalHamiltonian}
\end{equation}
where $I$ is the identity matrix and
\begin{equation}
\mathcal{M}=\begin{pmatrix}0&i(\hat{a}_{g}^{\dagger}-\hat{a}_{d})&0\\
-i(\hat{a}_{g}-\hat{a}_{d}^{\dagger})&0&(\hat{a}_{g}^{\dagger}-\hat{a}_{d})\\
0&-i(\hat{a}_{g}-\hat{a}_{d}^{\dagger})&0\end{pmatrix}.
\end{equation}

The units of the time, length and energy are $1/\omega$, $a_{\perp}=\sqrt{\hbar/M_{a}\omega}$ and $\hbar\omega$, respectively. The dimensionless spin-orbit coupling strength $\kappa=\gamma/\sqrt{\hbar\omega/M_{a}}$.
%In the wavefunction basis expanded by $|\chi_{n_{d},n_{g}}\rangle$, the single-particle Hamiltonian in Eq. (\ref{SingleParticalHamiltonian}) is diagonalized to find out the energy spectrum. In Fig. \ref{SPspectrumFIG}, the single-particle spectrums with spin-orbit coupling strength $\kappa=4,6$ are given. As the spin-orbit coupling strength increases, the spacing between energy levels with the same quantum number $n$ gets smaller.
In the wavefunction basis expanded by $|\chi_{n_{d},n_{g}}\rangle$, the single-particle Hamiltonian in Eq. (\ref{SingleParticalHamiltonian}) is diagonalized to find out the energy spectrum. The single-particle spectrum is shown in Fig. \ref{SPSpectrum}. It is shown that the spacing between energy levels with the same quantum number $n$ gets smaller as the spin-orbit coupling strength increases and the levels incline to resemble the well-known Landau levels spectrum structure.

\bigskip
\subsection{Weakly Interacting Bosons}

For a weakly interacting N-boson system, only the lowest Landau levels are occupied as long as $c_{0}N\ll\hbar\omega$. Thus, the field operator can be expanded as $\Psi=\sum_m\Psi_{m}\hat{a}_{m}$, where $\Psi_{m}$ is the single-particle states with $n=0$, and $\hat{a}_{m}$ is the corresponding annihilation operator. The many-body Hamiltonian can be rewritten in the second quantization form as
\begin{equation}
\hat{H}=\sum_{m}\epsilon_{m}\hat{a}^{\dagger}_{m}\hat{a}_{m}+\sum_{i,j,k,l}(U_{ijkl}+V_{ijkl})
\hat{a}^{\dagger}_{i}\hat{a}^{\dagger}_{j}\hat{a}_{k}\hat{a}_{l}, \label{ManybodyHamiltonian}
\end{equation}
where
\begin{eqnarray}
\!U_{ijkl}\!&=&\frac{1}{2}d_{0}\sum_{p,q}\int d\mathbf{r}(\psi_{p}^{i})^{*}(\psi_{q}^{j})^{*}(\psi_{q}^{k})(\psi_{p}^{l}),\\
\!V_{ijkl}\!&=&\!\frac{1}{2}d_{2}\!\sum_{p,q,r,s}\!\int d\mathbf{r}\!(\psi_{p}^{i})^{*}(\psi_{q}^{j})^{*}\!\mathbf{S}_{pr}\!\cdot\!\mathbf{S}_{qs}\!(\psi_{s}^{k})(\psi_{r}^{l}),
\end{eqnarray}
and $\epsilon_{m}$ is the single-particle energy corresponding to $\Psi_{m}$.

In $U_{ijkl}$ and $V_{ijkl}$, the summation subscript parameters have values $p,q,r,s=+, 0, -$ and $d_{0,2}=c_{0,2}/\sqrt{2\pi}a_{z}a_{\perp}^{2}\hbar\omega$. The many-body Hamiltonian can be solved numerically by using exact diagonalization in the Fock space. The numerical results from the exact diagonalization are also confirmed within the mean-field frame. In the mean-field frame, the operators $\hat{a}_{m}$ are replaced by complex numbers, and the mean-field energy are minimized under constrains of total particle number $N$ and/or total angular momentum $J_z$ \cite{LiuXiaoJi2012}.
\bigskip

\section{Fragmented State}\label{sec2}
\subsection{Fragmented Condensates Preserved by Total Angular Momentum Conservation}
In the following, we consider implicitly the Fock space spanned by states $|n_{0},n_{1},...,n_{k}\rangle$ with total particle number $N=\sum_{j=0}^{k}n_{j}$ and total angular momentum $J_{z}=\sum_{j=0}^{k}jn_{j}$ unless specified otherwise. Here, $n_{j}$ denotes the occupation of the single-particle state $\Psi_{m}$. In the calculation process, the convergence of the numerical results is required to ensure the numerical accuracy.

\begin{figure}[htbp]
\centering
\includegraphics[width=9.0cm]{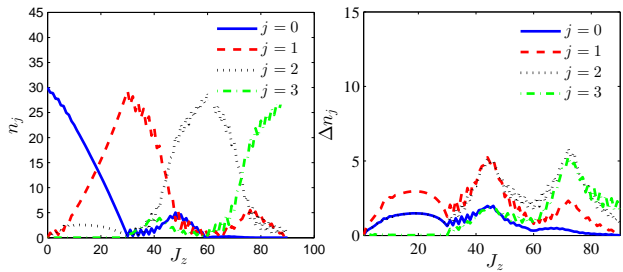}
\caption{Particle number distribution among the single-particle states denoted by $j$ with respect to the total angular momentum $J_z$ (a), and the corresponding particle number fluctuations $\Delta n_{j}$ (b). The spin-orbit coupling strength and the total particle number are $\kappa=4$ and $N=30$, respectively.}
\label{ParticalNumberAndFluctuation}
\end{figure}

As explained in Ref. \cite{XJLiu2001}, the one-body density matrix can be written as
\begin{equation}
n_{ij}=\langle \Psi_{N,J_{z}}|\hat{a}^{\dagger}_{i}\hat{a}_{j} |\Psi_{N,J_{z}}\rangle,
\end{equation}
where $|\Psi_{N,J_{z}}\rangle$ is the ground state of $N$-boson system with total angular momentum $J_{z}$. Because of the conservation of the total angular momentum, the one-body density matrix is diagonal, i.e. $n_{ij}=n_{i}\delta_{ij}$, indicating the eigenvalues are the occupation numbers of the single-particle states. The corresponding fluctuation of the occupation number is $\Delta n_{j}=\sqrt{<n_{j}^{2}>-<n_{j}>^{2}}$. In Fig. \ref{ParticalNumberAndFluctuation}, we plot the particle number distribution among the single-particle states and the fluctuations of the occupation numbers with respect to the total angular momentum $J_{z}$. As the total angular momentum increases, the particles choose single-particle states with higher characteristic angular momentum. The particle number distribution $n_{j}$ with respect to the total angular momentum $J_{z}$ is generally characterized by two peaks. Take $n_{0}$ for example, the second peak locates at $J_{z}/N=1.6$, indicating simultaneous particle occupations among single-particle states with $j=0$, $j=1$, and $j=2$.

\begin{figure}[htbp]
\centering
\includegraphics[width=9.0cm]{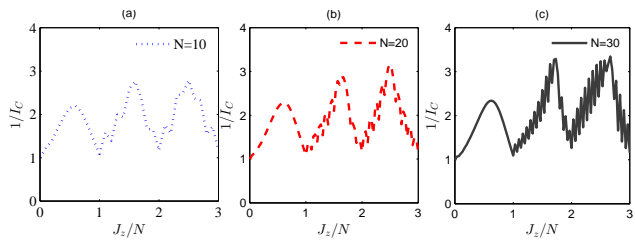}
\caption{The inverse participation ratio for a system of $N=10$ (a), $N=20$ (b) and $N=30$ (c), showing the fragmentation of the N-boson system is an universal characteristic regardless of the particle number.
The spin-orbit coupling strength is $\kappa=6$. }
\label{NumberChange}
\end{figure}

\begin{figure}[htbp]
\centering
\includegraphics[width=7.0cm]{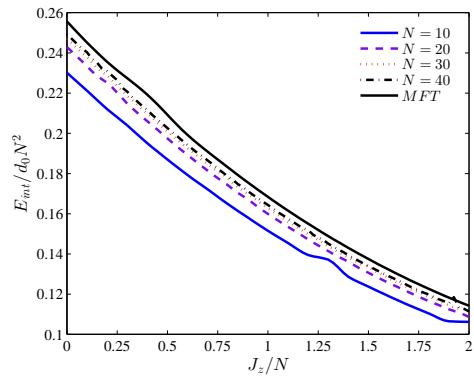}
\caption{The interaction energy with respect to the total angular momentum for Bose gas with $N=20,40,60$, respectively. For comparison, the interaction energy predicted by the mean-field theory is given (black solid line). As the total particle number $N$ increases, the interaction energy becomes closer to the mean-field interaction energy.}
\label{meanfield}
\end{figure}

The number of significantly occupied single-particle states can be described qualitatively by the inverse participation ratio $1/I_{C}$, where $I_{C}=\sum_{j}{\left(n_j/N\right)^2}$. To show that the existence of the fragmented condensation does not depend on the particle number, we plot the inverse participation ratio with respect to the total particle number $N$. As shown in Fig. \ref{NumberChange}, the overall profile of the inverse participation ratio resembles each other, exhibiting roughly periodical profiles. The peak value of the inverse participation ratio increases as the particle number gets larger. The global minima ($\simeq1$) appears at $J_{z}/N=0,1,2,3$, a signature of a single condensate. All the peak values are larger than $2$, indicating the fragmented characteristic of the corresponding ground state.

As found in scalar Bose gas \cite{XJLiu2001,Ho2000}, the fragmented ground state can be transformed into a single condensate in response
to an arbitrarily weak asymmetric perturbation. For the scalar Bose gas, the fragmented state has the same energy as that of the single condensate state in the limit of of $N\rightarrow\infty$. In the spin-orbit coupled Bose gas, similar phenomenon is observed. The interaction energy for boson systems with $N=10,20,30,40$ are plotted in Fig. \ref{meanfield}. As the total particle number $N$ increases, the interaction energy becomes closer to the mean-field interaction energy \cite{appendix}, indicating that the fragmented state inclines to break into a single condensate state.

\subsection{Effects of the Spin-orbit Coupling Strength and Spin-dependent Interactions}

\begin{figure}[htbp]
\centering
\includegraphics[width=7.0cm]{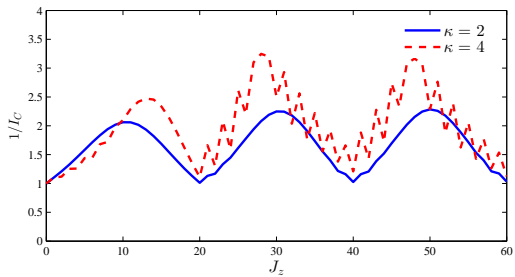}
\caption{The inverse participation ratio, which indicates qualitatively the number of significantly occupied single-particle states with spin-orbit coupling strength $\kappa=2$ and $\kappa=4$, respectively. For stronger spin-orbit coupling, the value of the inverse participation ratio and its oscillation is larger.}
\label{fig3}
\end{figure}

As shown in Sec. \ref{sec1}, the increase of the spin-orbit coupling strength changes the energy spectrum evidently. In this section, the effect of the spin-orbit coupling strength on the particle number distribution and the inverse participation ratio is analyzed.
The variation of the spin-orbit coupling strength influences the particle distribution. Figure \ref{fig3} shows the approximately periodical inverse participation ratio for $\kappa=2$ and $\kappa=4$. For spin-orbit coupling strength $\kappa=2$, the value of $1/I_{C}$ is generally smaller than that when $\kappa=4$. The oscillation amplitude becomes relatively larger when the spin-orbit coupling strength is stronger. It can be interpreted that fragmented ground condensates are more favorable for higher angular momentums with strong spin-orbit coupling. This should be explained by resorting to the single-particle energy spectrum. As the spin-orbit coupling strength gets stronger, the energy levels tend to resemble the well-known Landau levels, and the spacing between energy levels with the same quantum number $n$ gets smaller. Thus for the same interaction strength, it is easier for particles to populate into higher $m$ single-particle states, resulting in the higher fragmentation of the bosons.

The spin-dependent interaction determines the ground state phase of the Bose gas. For $^{87}$Rb Bose gas, the spin-dependent interaction term is minus, thus the ground state is ferromagnetic. For $^{23}$Na Bose gase, however, the ground state is antiferromagnetic. The variation of the spin-dependent interaction affects the particle number distribution and further the inverse participation ratio, changing the fragmentation characteristic of the Bose gas.

\begin{figure}[htbp]
\centering
\includegraphics[width=9.0cm]{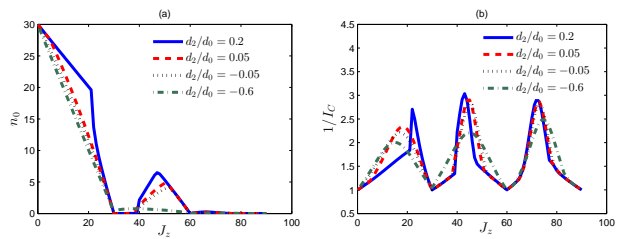}
\caption{The particle number of the single-particle state with $j=0$ and the inverse participation ratio. The total particle number is $N=30$. As the Bose gas renders from anti-ferromagnetic to ferromagnetic, the inverse participation ratio inclines to be suppressed.}
\label{c2Change}
\end{figure}

In Fig. \ref{c2Change} (a), the second peak value of $n_0$ becomes lower as the spin-dependent interaction decreases. At the same time, the slope of the $n_0$ with respect to the total angular momentum $J_z$ gets larger in the range of $J_z/N<1$. This means that the particle redistributes toward the center value of the angular momentum $J_z$ from both ends. As $d_2/d_0$ reaches $0.6$ the second peak in $n_0$ nearly disappears. This particle number distribution can also reflected by the inverse participation ratio. As shown in Fig. \ref{c2Change} (b), all the three peak values decrease as the spin-dependent interaction changes in the process from antiferromagnetic to ferromagnetic. Thus, we conclude that the fragmented state is more evident in anti-ferromagnetic Bose gas than that in ferromagnetic one.

\bigskip

\begin{figure}[htbp]
\centering
\includegraphics[width=8.5cm]{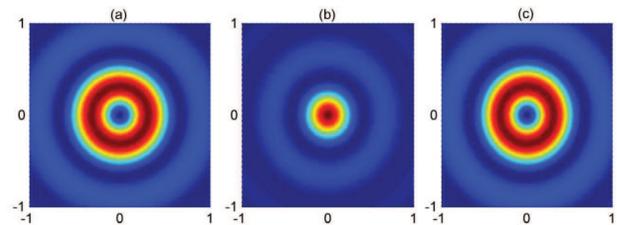}
\caption{The density profile of components with magnetic quantum number $m_F=1$ for (a), $m_F=0$ for (b), $m_F=-1$ for (c). The interaction parameter is $d_{0}=0.005$. The total particle number and the spin-orbit coupling strength are $N=30$ and $\kappa=4$, respectively. The total angular momentum is $J_z/N=0$.}
\label{Lattice_1}
\end{figure}

\section{Ground State with Zero or Fractional Total Angular Momentum}\label{sec3}
Similar to the results in \cite{XJLiu2001}, the fragmented state and its corresponding single condensate state have the same energies in the limit of $N\rightarrow\infty$ as shown in Fig. \ref{meanfield}. Thus, even a perturbation of order O(1/N) can be strong enough to take the fragmented state into a single condensate state. Consequently, it is reasonable to expect a single condensate ground state. To find the single condensate ground state of the Hamiltonian $\hat{H}= \hat{H}_0+ \hat{H}_{int}$ without the confinement of specified angular momentum, the single-particle states with both $m\geq0$ and $m<0$ have to be considered simultaneously. As found in Ref. \cite{HuiHu2012}, the ground state chooses a spontaneous total angular momentum with $J_{z}=\pm1/2$ or $J_{z}=0$, which is determined by the interaction parameter. For the spin-orbit coupled spin-1 Bose gas, the ground state is found with zero total angular momentum $J_{z}$ when the spin-orbit coupling is weak with an interaction range from $Nd_{0}=0-0.6$ (the spin-dependent interaction is specified a hundredth of the spin-independent interaction, which is consistent with experimental parameters). In the strong spin-orbit coupling limit, however, either zero or finite total angular momentum ground state is found, which is determined by the interaction strength.

\begin{figure}[t]
\begin{minipage}[b]{0.5\textwidth}
\centering
\includegraphics[width=8.5cm]{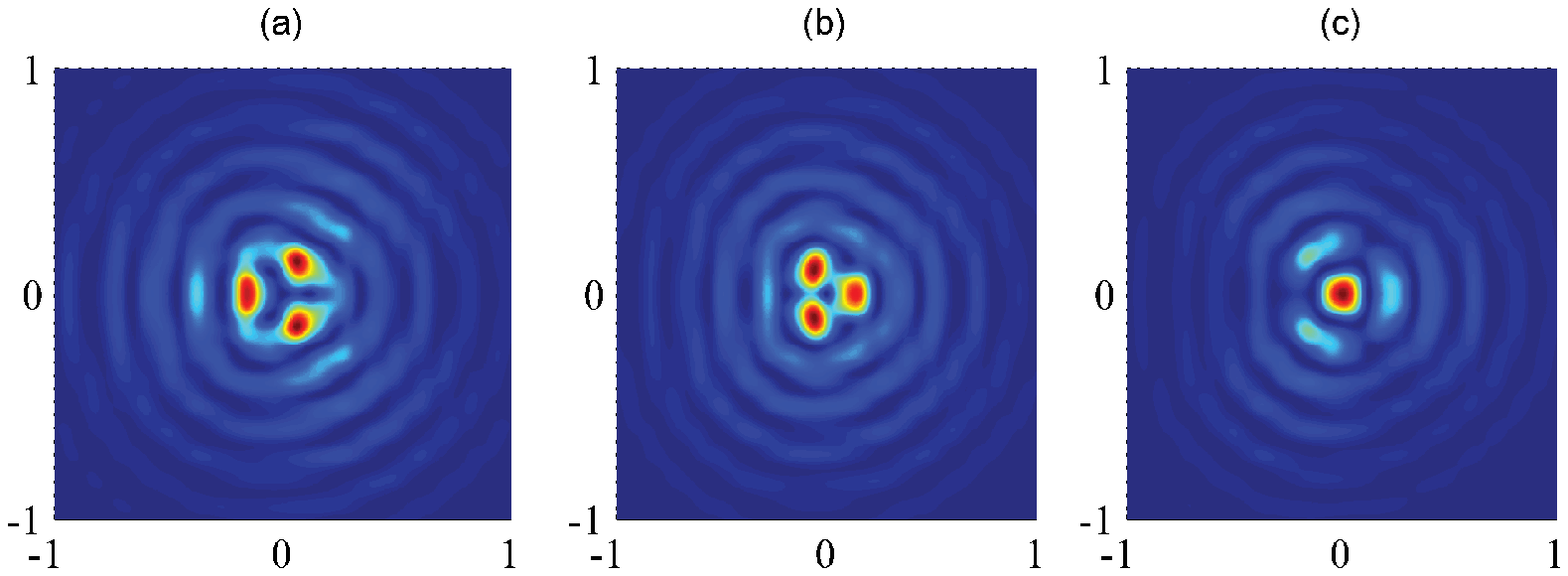}
\end{minipage}
\begin{minipage}[b]{0.5\textwidth}
\centering
\includegraphics[width=8.5cm]{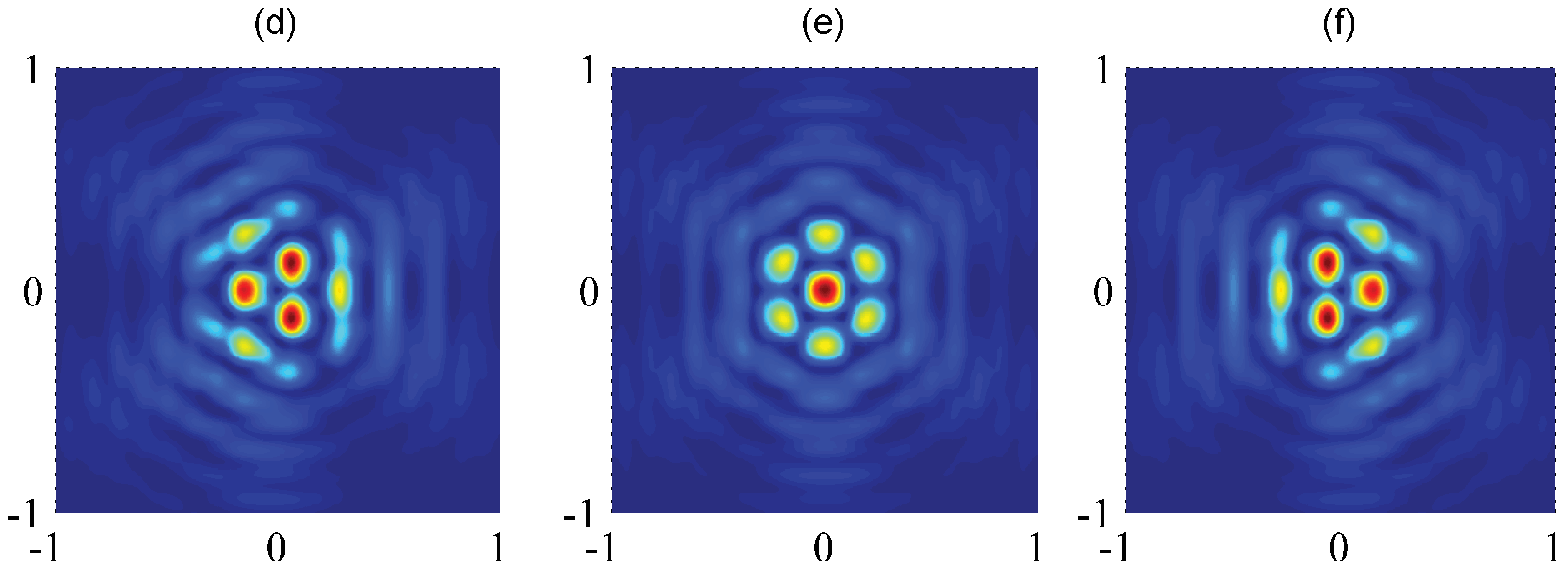}
\end{minipage}
\caption{The density profile of components with magnetic quantum number $m_F=1$ for (a) and (d), $m_F=0$ for (b) and (e), $m_F=-1$ for (c) and (f). The interaction parameter is $d_{0}=0.005$ for (a)-(c), and $d_{0}=0.01$ for (d)-(f). The total particle number and the spin-orbit coupling strength are $N=30$ and $\kappa=12$, respectively. For (a)-(c), the total angular momentum is $J_z/N=1/3$ and it is $J_z/N=0$ for (d)-(f).}
\label{Lattice_2}
\end{figure}

In Fig. \ref{Lattice_1}, the ground state density distributions with $\kappa=4$, $Nd_{0}=0.15$ are given \cite{MYexplanation}. It is numerically found that this ground state configuration is not sensitive to the relatively weak interaction. The density profiles are nearly the same for the case of $Nd_{0}=0.3$, and both of them carry a zero total angular momentum $J_{z}/N=0$. In fact, the single-particle state with $j=0$ is mainly occupied due to considerable energy spacings between levels with $j=0,\pm1,...$ within the interaction range $Nd_{0}=0-0.6$ and the probability distribution among single-particle states is symmetric. For strong spin-orbit coupling, there is only the single-particle state with $j=0$ occupied when the interaction is absent. As the interaction comes into presence and gets stronger, however, the probability distribution among single-particle states becomes asymmetric and the ground state carries $J_z/N=1/3$ angular momentum. The density profile of the ground state with $Nd_{0}=0.15$ is shown in Fig. \ref{Lattice_2} (a)-(c), which carries  a finite angular momentum. As the interaction strength increase further, hexagonal lattice ground state carrying zero angular momentum is observed as shown in Fig. \ref{Lattice_2} (d)-(f).

In this paragraph, we present the experimental relevance, taking $^{87}$Rb and $^{23}$Na for example. The two-dimensionality of the present system can be routinely realized by imposing a strong harmonic confinement with $\omega_z\gg\omega$. The interaction parameters can be generally tuned by varying the atom species, trapping frequencies. Referring to that in Refs. \cite{TinLunHo,Song2012}, the spin-independent and spin-dependent parameters read $d_0=0.0647$, $d_2=-0.0006$ for $^{87}$Rb gas and $d_0=0.0155$, $d_2=0.0006$ for $^{23}$Na gas if $\omega_z=600\times2\pi$ and $\omega=20\times2\pi$ are set experimentally.  Considering the experimental setups in Refs. \cite{YJLin2009,PWang2012}, the dimensionless spin-orbit coupling strength parameter $\kappa={\gamma}/{\sqrt{\hbar\omega/M_{a}}}=\frac{2\pi\hbar}{M_{a}\omega}\mathrm{sin}(\theta/2)/\sqrt{{\hbar\omega}/{M_{a}}}$. To obtain dimensionless spin-orbit coupling strength values $\kappa=2,4,12$ for Raman beams with wavelength $\lambda=800nm$, the angle $\theta$ between two Raman beams should be chosen as $\theta=6.23^{\circ},  12.47^{\circ}, 38.03^{\circ}$ for $^{23}$Na gas. Because the atom mass is different for $^{87}$Rb, the corresponding angle $\theta$ chooses $\theta=9.32^{\circ},  18.71^{\circ}, 58.37^{\circ}$.

\bigskip

\bigskip
\section{CONCLUSIONS} \label{sec4}
We mainly explore spin-orbit coupled spin-1 Bose gas with a weak interaction in external harmonic trap by the exact diagonalization and mean-field theory. The inverse participation ratio shows the appearance of the fragmented condensate state. The particle number distribution among the single-particle states is affected evidently by the spin-orbit coupling strength and the spin-dependent interaction. Large spin-orbit coupling strength raises the peaks of the inverse participation ratio line, manifesting the fragmented condensate signature. Similar to the scalar Bose gas, the interaction energy of the fragmented state approaches to the mean-field interaction energy of the single condensate state as the total particle number $N\rightarrow\infty$. Without the confinement of the specified angular momentum, the ground state favors a zero ($J_{z}/N=0$) or finite ($J_{z}/N=\pm1/3$) total angular momentum, which is determined by both the interaction and the spin-orbit coupling strength.

\textbf{Acknowledgments}
{This work was supported by the NKBRSFC under grants Nos. 2011CB921502, 2012CB821305, NSFC under grants Nos. 61227902, 61378017, 11311120053. }


\begin{thebibliography}{70}
\bibitem{appendix} The field operators in interaciton Hamiltonian of Eq. (\ref{Hamiltonian2}) are replaced by their mean-field complex order parameters. Then the complex order parameters are expanded in terms of the single particle states in Eq. (\ref{SingleParticleState}). The interaction energy is minimized to determine the variational parameters.
\end{thebibliography}
\end{document}